\documentclass[conference, 9pt]{IEEEtran}
\IEEEoverridecommandlockouts

\usepackage{cite}
\usepackage{amsmath,amssymb,amsfonts}
\usepackage{algorithmic}
\usepackage{graphicx}
\usepackage{textcomp}
\usepackage{xcolor}
\usepackage{amsmath,graphicx}
\usepackage{hyperref}       
\usepackage{url}            
\usepackage{booktabs}       
\usepackage{amsfonts}       
\usepackage{nicefrac}       
\usepackage{microtype}      
\usepackage{lipsum}
\usepackage{graphicx}
\usepackage{amsmath}
\usepackage{caption}
\usepackage{subcaption}
\usepackage{multirow}
\graphicspath{ {./images/} }
\usepackage{amsfonts}       
\usepackage{color, colortbl, arydshln}
 
\newcommand{\myparagraph}[1]{\noindent\textbf{#1} \hspace{.15cm}} 
\def\BibTeX{{\rm B\kern-.05em{\sc i\kern-.025em b}\kern-.08em
    T\kern-.1667em\lower.7ex\hbox{E}\kern-.125emX}}

\begin{document}

\title{AnCoGen: Analysis, Control and Generation of Speech\\ with a Masked Autoencoder\\
\thanks{This research was partly supported by the French National Research Agency (ANR) as a part of the DEGREASE project (ANR-23-CE23-0009).
}}

\author{\IEEEauthorblockN{Samir Sadok\IEEEauthorrefmark{1}, Simon Leglaive\IEEEauthorrefmark{2}, Laurent Girin\IEEEauthorrefmark{3},  Gaël Richard\IEEEauthorrefmark{4}, Xavier Alameda-Pineda\IEEEauthorrefmark{1}}
\IEEEauthorblockA{\IEEEauthorrefmark{1}Inria at Univ. Grenoble Alpes, CNRS, LJK, France}
\IEEEauthorblockA{\IEEEauthorrefmark{2}CentraleSup\'elec, IETR UMR CNRS 6164, France}
\IEEEauthorblockA{\IEEEauthorrefmark{3}Univ. Grenoble Alpes, CNRS, Grenoble INP, GIPSA-lab, France}
\IEEEauthorblockA{\IEEEauthorrefmark{4}LTCI, Télécom Paris, Institut polytechnique de Paris, France}
}


\maketitle

\begin{abstract}
This article introduces AnCoGen, a novel method that leverages a masked autoencoder to unify the analysis, control, and generation of speech signals within a single model. AnCoGen can analyze speech by estimating key attributes, such as speaker identity, pitch, content, loudness, signal-to-noise ratio, and clarity index. In addition, it can generate speech from these attributes and allow precise control of the synthesized speech by modifying them. Extensive experiments demonstrated the effectiveness of AnCoGen across speech analysis-resynthesis, pitch estimation, pitch modification, and speech enhancement. 
Code and audio examples are available online\footnote{\label{note1}\url{https://samsad35.github.io/site-ancogen}}.

\end{abstract}
\section{Introduction}

Over the years, many speech processing algorithms have been developed to analyze, transform, and synthesize speech signals. This includes time stretching, pitch shifting, timbre modification, and speech enhancement in noise. Conventional parametric approaches rely on a signal model whose parameters are estimated during the analysis stage and explicitly modified in the transformation and synthesis stages. Examples of well-known parametric signal models include linear predictive coding \cite{markel1976linear}, 
sinusoidal models \cite{mcaulay1986speech, george1997speech} and harmonic-plus-noise models \cite{serra1990spectral,laroche1993hns, richard1996analysis}.
Non-parametric methods 
do not require the explicit estimation and manipulation of speech parameters. Historical examples of such methods for pitch- and time-scale modification include PSOLA \cite{moulines1990pitch} and the phase vocoder \cite{laroche1999improved}.
More recently, vocoders such as STRAIGHT 
\cite{kawahara2006straight} and WORLD \cite{morise2016world} were proposed for real-time manipulation of speech. They can be seen as learning-free encoding/decoding-based methods in which signal processing techniques are used to decompose (resp. synthesize) a speech signal into (resp. from) a set of acoustic features consisting of the fundamental frequency ($f_0$), spectral envelope, and aperiodicity. 


Today, the most effective methods for manipulating speech signals are based on deep learning. Various neural encoder-decoder models have been proposed to encode a speech signal into a compact representation that can then be decoded with minimal loss of quality and intelligibility. Neural audio codecs such as SoundStream \cite{zeghidour2021soundstream} and EnCodec \cite{defossez2023high} focus on extracting low-bitrate discrete units while achieving high-quality reconstruction. Encoders based on self-supervised learning (SSL), such as HuBERT \cite{hsu2021hubert}, have also been used to extract a representation of the linguistic content of speech, which can then be combined with acoustic attributes (e.g., $f_0$) and speaker identity to reconstruct the speech signal \cite{polyak2021speech}. Existing neural encoder-decoder models achieve high-quality speech reconstruction but significantly lack controllability for speech signal manipulation and robustness to noise and reverberation.

\begin{figure}[t!]
    \centering
    \includegraphics[width=0.9\linewidth]{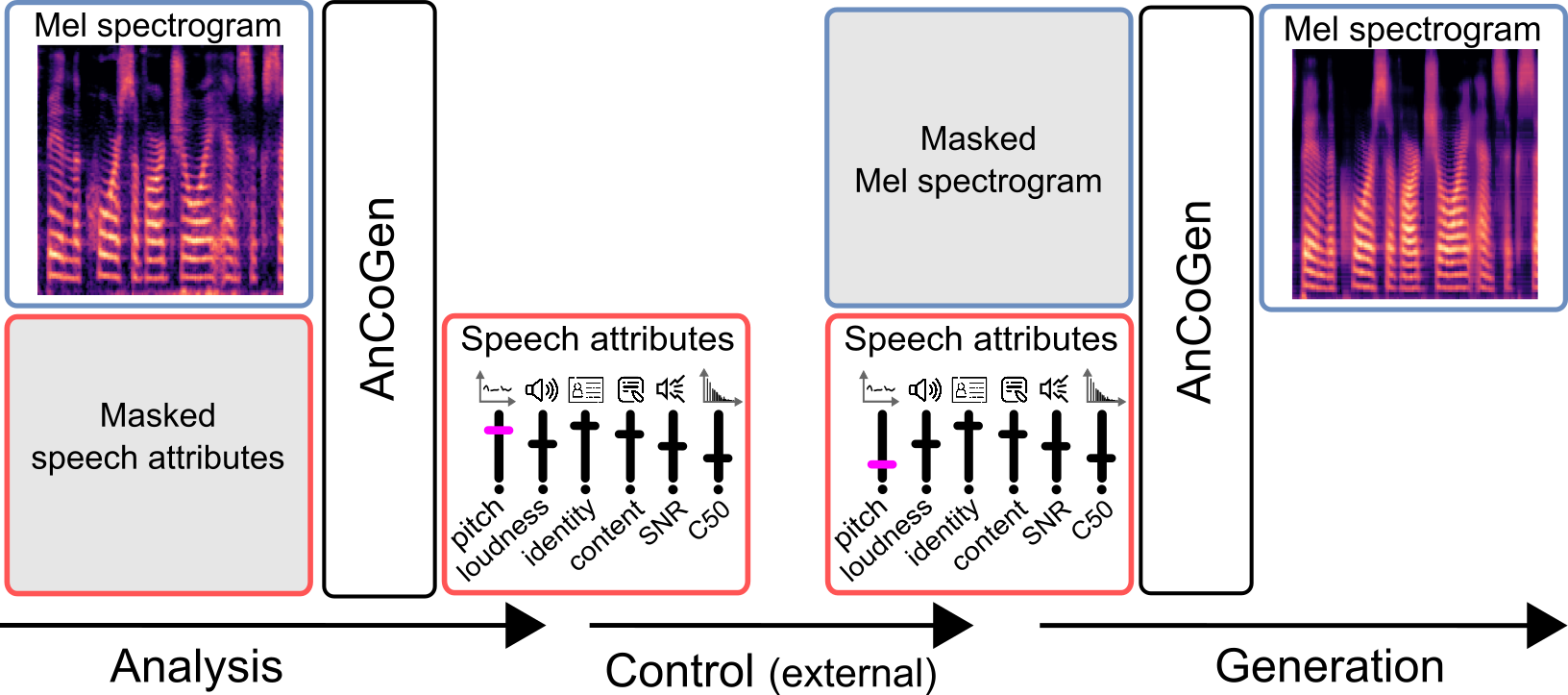}
    \caption{Analysis, control, and generation of speech with AnCoGen.}
    \label{fig:AnCoGen_high_level}
\end{figure}

In this paper, we introduce AnCoGen for analyzing, controlling, and generating speech signals. AnCoGen can decompose a speech signal into a compact but comprehensive set of speech attributes representing not only the linguistic content, prosody (pitch and loudness), and speaker identity, but also the acoustic recording conditions in terms of noise and reverberation. As in the previously proposed neural architectures for speech manipulation,  the encoded attributes can be controlled before resynthesis, enabling various applications, including pitch transformation, voice conversion, noise suppression, and dereverberation. However, in the previous works, separate encoder and decoder models were designed to estimate the speech attributes and synthesize the speech signal, respectively. With AnCoGen, we propose a fundamentally different approach. We draw inspiration from the masked autoencoder (MAE) model \cite{he2022masked} to learn a bidirectional mapping between a Mel-spectrogram and the speech attributes, hence leading to \emph{a single model for the analysis and the resynthesis}, as illustrated in Fig.~\ref{fig:AnCoGen_high_level}.  The Mel-spectrogram and the attributes can be seen as two different representations of the 
uttered speech. The Mel-spectrogram is a low-level representation close to the raw signal, while the attributes are a higher-level representation more suitable for controlling and manipulating speech. Following the learning strategy of the multimodal MAE \cite{sadok2023avector}, training the AnCoGen model consists of reconstructing masked elements in a given representation from visible elements in this same representation and the other. In doing so, the model learns the inter- and intra-representation dependencies. Then, by masking entirely one representation (e.g., the Mel-spectrogram), it is possible to generate the other (e.g., the speech attributes), and this generative process operates in both ways, with the same MAE model (see Fig.~\ref{fig:AnCoGen_high_level}). Combined with a HiFi-GAN neural vocoder \cite{kong2020hifi}, the proposed approach allows us to analyze, transform, and synthesize speech signals efficiently with a single model.

\begin{figure*}[!ht]
\centering
\includegraphics[width=.8\linewidth]{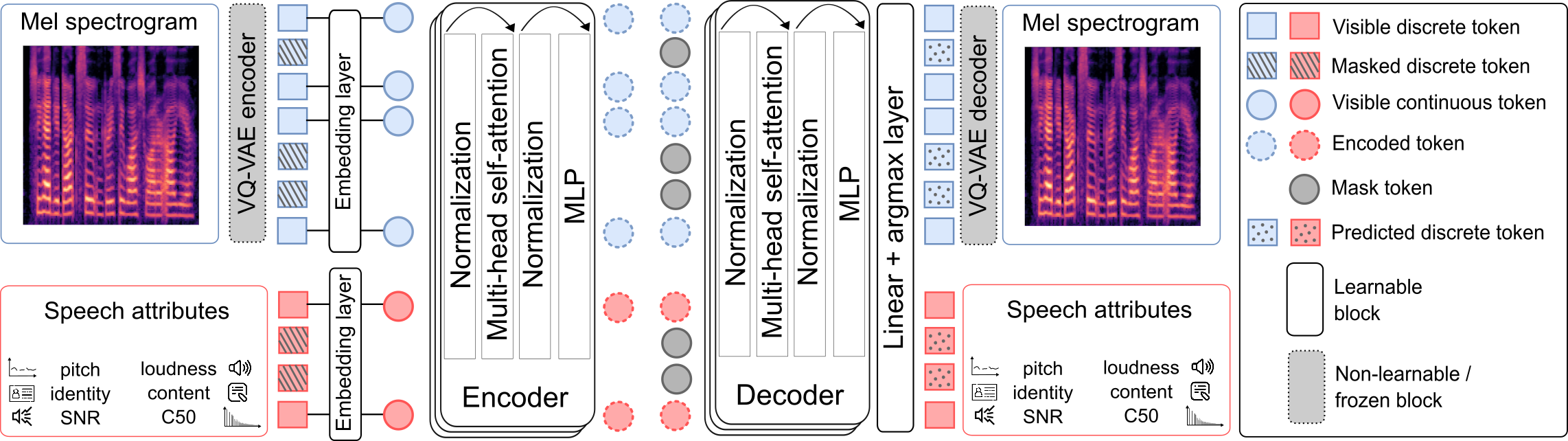}
\caption{Overall architecture of AnCoGen, to read from left (input) to right (output).}
\label{fig:AnCoGen_archi}
\end{figure*}

\section{The proposed model: AnCoGen}

We first provide a general overview of the proposed AnCoGen model, with technical details in the following subsections. The model, represented in Fig.~\ref{fig:AnCoGen_archi}, is based on an encoder-decoder Transformer model that auto-encodes the concatenation of a speech Mel-spectrogram (MS) and corresponding speech attributes (SAs). The general principle is that, at training time, these two speech representations are partially masked to learn their intra- and inter-dependencies, whereas at inference time, the masking depends on the task: For speech analysis (resp. generation), the SAs (resp. MS) are not available, and they are thus totally masked at the input and estimated by the MS-to-SA (resp. SA-to-MS) mapping. For generation, the speech waveform is reconstructed from the MS using the pre-trained HiFi-GAN neural vocoder~\cite{kong2020hifi} (not shown in Fig.~\ref{fig:AnCoGen_archi}).

In a few more details, the general pipeline is as follows: The MS and SAs are first separately quantized into a sequence of discrete indices called \textit{discrete tokens} (blue and red squares in Fig.~\ref{fig:AnCoGen_archi}), which are then (partially) masked. The sequence of remaining visible discrete tokens is converted by the embedding layer into a sequence of real-valued vectors, called \textit{continuous tokens}, taken from a learned codebook (blue and red circles). The continuous tokens are subsequently passed to the Transformer encoder. The resulting sequence of encoded continuous tokens is completed with learned \textit{mask tokens} (grey circles) before being processed by the Transformer decoder. Finally, a linear + argmax layer outputs a sequence of decoded discrete tokens, which can be used to decode an MS and/or SAs. The model (encoder, decoder, codebooks and mask token) is trained by computing the cross-entropy loss between the predicted and ground-truth discrete tokens.

\subsection{Speech representations}
\label{subsec:speech_rep}

The MS computation is detailed in Section~\ref{subsec-exp-setup}. The SAs are the following six high-level speech features.
$A_1$: The output embedding vector sequence of the pre-trained HuBERT model \cite{hsu2021hubert}, which is assumed to represent the speech signal linguistic content; $A_2$: The pitch contour $f_0$ (in~Hz), as estimated by the pre-trained CREPE model \cite{kim2018crepe} (we used the PyTorch implementation \texttt{torchcrepe}\footnote{\url{https://github.com/maxrmorrison/torchcrepe}}); $A_3$: The loudness, here defined as the root mean square (RMS) signal level measured over short-time overlapping windows; $A_4$: The speaker identity, represented by the embedding vector provided by the pre-trained model of \cite{desplanques2020ecapa}; $A_5$: The signal-to-noise ratio (SNR; in dB), as estimated from overlapping frames of the input signal by the pre-trained Brouhaha model \cite{lavechin2023brouhaha}; $A_6$: The clarity index at 50~ms (C50; in dB), defined as the energy ratio of the early and late parts of a room impulse response, also estimated frame-wise by Brouhaha \cite{lavechin2023brouhaha}.
Note that when generating the training dataset, all the SAs are estimated using off-the-shelf techniques (based either on signal processing or on deep learning). When using noisy and reverberant speech signals for training AnCoGen, we assume the availability of the corresponding clean speech signals to estimate attributes $A_1$ to $A_4$.

\subsection{Token representations}
\label{subsec:tokens}

The MS discrete tokens are obtained from the discrete latent representation of a pre-trained and frozen vector-quantized variational autoencoder (VQ-VAE) \cite{van2017neural}. Following \cite{sadok2023vector, sadok2023avector}, the VQ-VAE operates independently on each time frame and it is fully convolutional on the Mel-frequency axis, thus preserving the time-frequency structure of the original MS representation. For each time frame, the VQ-VAE encoder outputs a vector $\mathbf{x}_t \in \mathbb{X} = \{1,..., C\}^F$ of discrete indices (corresponding to the set of $F$ quantized latent vectors in the VQ-VAE codebooks, each codebook being of size $C$). $\mathbf{x}_t$ is a blue square in Fig.~\ref{fig:AnCoGen_archi} and a (time) sequence of $N$ MS discrete tokens $\mathbf{x} =\left\{\mathbf{x}_t \in \mathbb{X}  \right\}_{t=1}^{N}$ is the vertical set of blue squares. We also experimented using tokens created from 2D (time-frequency) patches of indices, but the frame-based approach led to better performance. A sequence of MS discrete tokens can be converted back to an MS using the VQ-VAE decoder.

The SA discrete tokens result from the quantization of the six SAs $A_1$--$A_6$. We denote by $\mathbf{a}_i = \left\{\mathbf{a}_{i,t} \in \mathbb{A}_i \right\}_{t=1}^{M_i}$ the (time) sequence of $M_i$ discrete tokens for $A_i$, $i \in \{1,...,6\}$, where one $A_i$ token corresponds to $D_i$ integer values between $1$ and $K_i$, i.e., $\mathbb{A}_i =\{1,..., K_i\}^{D_i}$. Note that in Fig.~\ref{fig:AnCoGen_archi}, due to space limitations, we represented with red squares one single sequence $\mathbf{a}_i$ for some index $i$, but we actually have 6 sequences of SAs at the input of the model.
Attributes $A_2$ ($f_0$), $A_3$ (loudness), $A_5$ (SNR), and $A_6$ (C50) are 1D real-valued sequences that are normalized, resampled (if needed to have the same length), and rounded to obtain sequences of discrete integer values. Each token $\mathbf{a}_{i,t}$ for $i \in \{2,3,5,6\}$ is then obtained by grouping $D_i$ contiguous discrete values from the corresponding sequence. Attributes $A_1$ (HuBERT representation) and $A_4$ (speaker identity embedding vector) are quantized using the k-means algorithm, as in \cite{van2022comparison}. The number of clusters is set to $K_1 = 100$ for $A_1$ and to the number of speakers in the training dataset for $K_4$. HuBERT already provides a sequential representation, while for the speaker identity attribute, a sequence of discrete tokens is obtained by repeating the cluster index. In practice, we have $(D_i, K_i) = (2, 100)$, $(4, 400)$, $(4, 100)$, $(4, 251)$, $(4, 128)$, and $(4, 128)$, for $i = 1, 2, 3, 4, 5, 6$. 

The MS and SA discrete tokens are transformed by the embedding layers into MS and SA continuous tokens, respectively, before being fed to the encoder. 
This means that each entry of a discrete token $\mathbf{x}_t$ or $\mathbf{a}_{i,t}$ is replaced by a real-valued vector taken in a learnable codebook of fixed size (each SA token has its own codebook), and the vectors for all the entries of a given token are concatenated to form a continuous token. The vector dimensions are chosen such that the resulting MS and SA continuous tokens are of the same size, since they are fed to the same encoder model as shown in Fig.~\ref{fig:AnCoGen_archi}. At the output of the encoder, the encoded continuous tokens sequence is completed with a learned mask token before being sent to the decoder. The decoder output is a ``full'' sequence of decoded/predicted discrete tokens that can be used for MS and SAs estimation (using the VQ-VAE decoder and inverse quantization, respectively). 

\subsection{Masking strategy}
\label{subsec:masking}

Training AnCoGen is achieved by randomly masking the input sequences of MS and SA tokens. The masking ratio (i.e., the proportion of tokens that are masked) for these two speech representations is drawn randomly according to a uniform distribution on the 1-simplex. This means that if $p \times 100~\%$ of the MS tokens are masked, then $(1-p) \times 100~\%$ of the SA tokens are masked, where $p$ is distributed uniformly between 0 and 1. Note that the masked tokens are chosen randomly and independently for the different SAs $A_i$, even if the masking ratio is the same. The training process involves two consecutive phases: First, the above-described ``coupled" masking strategy is applied, which enables to learn the inter- and intra-representation dependencies. Then an ``all or nothing" masking strategy is applied, in which one of the two speech representations (MS or SAs) is entirely masked out and the other remains entirely visible (i.e., $p=0$ or $1$). This helps AnCoGen to learn to estimate the SAs from an MS (analysis) and conversely (synthesis). At inference (analysis or synthesis), of course, only the ``all or nothing" strategy is applied.

\subsection{Encoder-decoder architecture}
\label{subsec:archi}

The encoder and decoder of AnCoGen each contain 12 multi-head self-attention (MHSA) blocks from the Vision Transformer (ViT) \cite{dosovitskiy2020image}. Each block comprises an MHSA module with 4 heads and a multilayer perceptron (MLP) module, with layer normalization preceding and residual connections following every module. This block is inspired by the attention layer of the original Transformer \cite{vaswani2017attention}. Position embeddings are added to the continuous tokens at the input of the encoder and decoder.

\section{Experiments}

\begin{table}[t]
\setlength{\tabcolsep}{4pt}
\centering
\resizebox{1.0\linewidth}{!}{ 
\begin{tabular}{ccccccccc}
\toprule
& PESQ \footnotesize{$\uparrow$} & STOI \footnotesize{$\uparrow$} & SI-SDR \footnotesize{$\uparrow$} & N-MOS \footnotesize{$\uparrow$} & WER \footnotesize{$\downarrow$} & COS \footnotesize{$\uparrow$}\\
\cmidrule{2-7}
GT waveform (seen) & $4.64$ & $1.00$ & $\infty$ & 4.50 & 0.57 & 1.00\\
GT waveform (unseen) & $4.64$ & $1.00$ & $\infty$ & 4.51 & 2.19 & 1.00\\
GT MS (seen) & 2.60 & 0.93 & -25.49 & 4.44 & 2.28 & 0.96 \\ 
AnCoGen (seen) & 1.57 & 0.82 & -17.43 & 4.41 & 3.03 & 0.90\\ 
GT MS (unseen) & 2.62 & 0.93 & -28.44 & 4.43 & 2.31 & 0.95 \\
AnCoGen (unseen) & 1.40 & 0.79 & -22.02 & 4.35 & 5.30 & 0.75\\
\bottomrule
\end{tabular}
}
\captionof{table}{Analysis-resynthesis results.}
\label{tab:quality-reconstruction}
\end{table}

\begin{figure}[t]
    \centering
    \includegraphics[width=\linewidth]{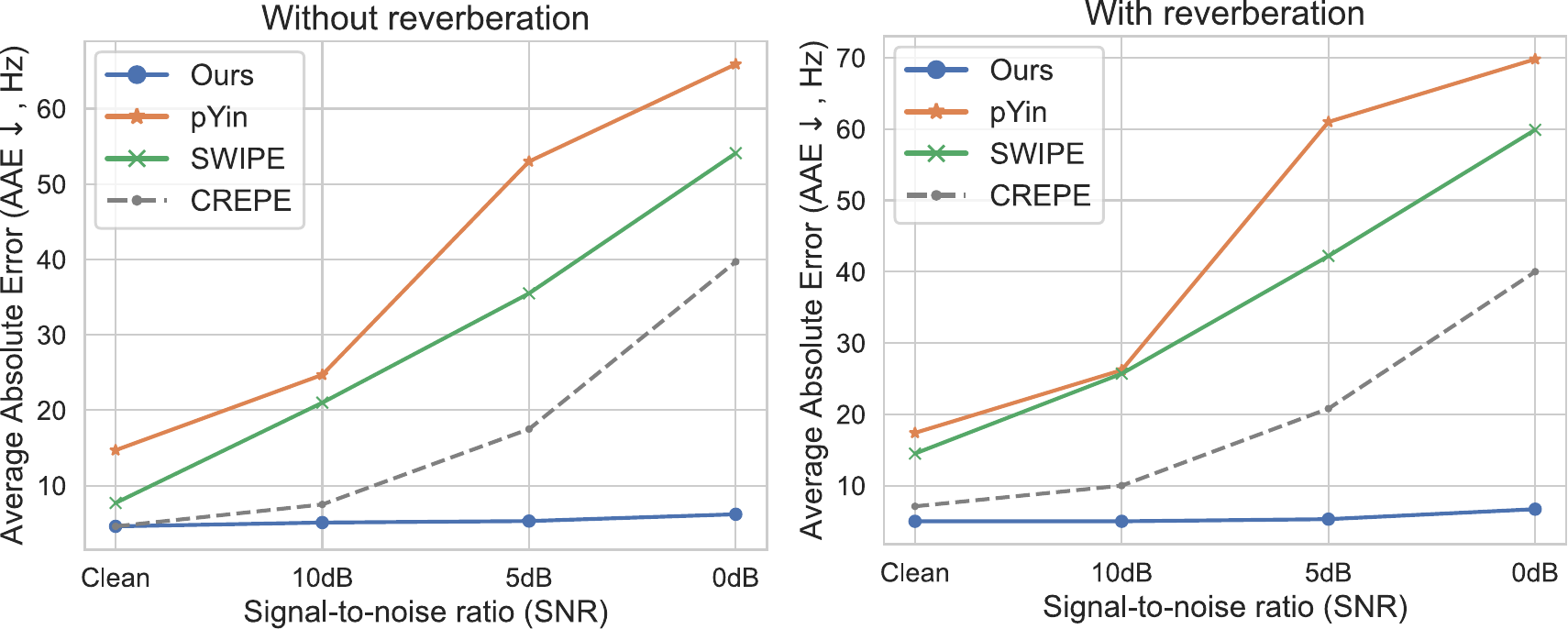}
    \caption{$f_0$ estimation results.}
    \label{fig:pitch_estimation}
\end{figure}

\begin{table}[t]
\setlength{\tabcolsep}{3pt}
\centering
\resizebox{1.0\linewidth}{!}{ 
\begin{tabular}{cccccccccc}
\toprule
\multicolumn{2}{c}{} & \multicolumn{2}{c}{+50 \%} & \multicolumn{2}{c}{+10 \%} & \multicolumn{2}{c}{-10 \%} & \multicolumn{2}{c}{-50 \%}\\ 
\cmidrule(lr){3-4} \cmidrule(lr){5-6} \cmidrule(lr){7-8} \cmidrule(lr){9-10}
 \multicolumn{2}{c}{}  & AAE \footnotesize{$\downarrow$} & N-MOS \footnotesize{$\uparrow$}  & AAE \footnotesize{$\downarrow$} & N-MOS \footnotesize{$\uparrow$} & AAE \footnotesize{$\downarrow$} & N-MOS \footnotesize{$\uparrow$} & AAE \footnotesize{$\downarrow$} & N-MOS \footnotesize{$\uparrow$} \\ \midrule
\multicolumn{2}{c}{WORLD \cite{morise2016world}}  & 8.20 & 4.03 & 8.92 & 4.11 & 6.19  & 3.87 & \underline{6.92} & 3.37  \\
\multicolumn{2}{c}{TD-PSOLA \cite{moulines1990pitch}}  & \underline{6.88} & \textbf{4.28} & \underline{6.56} & \textbf{4.45} & \underline{5.69}  & \textbf{4.43} & 13.10 & \underline{3.80}  \\ 
\multicolumn{2}{c}{AnCoGen (ours)} & \textbf{5.91} & \underline{4.27} & \textbf{5.54} & \underline{4.42} & \textbf{5.36} & \underline{4.37} & \textbf{5.82} & \textbf{4.30} \\ 
\bottomrule
\end{tabular}
}
\captionof{table}{Pitch shifting results (best score in each column is in bold, second best score is underlined).}
\label{tab:pitch-manipulation}
\vspace{-.25cm}
\end{table}

\begin{figure*}[t]
\centering
    \centering
    \resizebox{.82\linewidth}{!}{ 
    \begin{tabular}{cccccccccccccc}
    \toprule
         & \multicolumn{6}{c}{Matched conditions} & \multicolumn{6}{c}{Mismatched conditions} \\
        \cmidrule(lr){2-7} \cmidrule(lr){8-14}
         & N-MOS \footnotesize{$\uparrow$} & SIG \footnotesize{$\uparrow$} & BAK \footnotesize{$\uparrow$} & OVRL \footnotesize{$\uparrow$} & WER \footnotesize{$\downarrow$} & COS \footnotesize{$\uparrow$} & N-MOS \footnotesize{$\uparrow$} & SIG \footnotesize{$\uparrow$} & BAK \footnotesize{$\uparrow$} & OVRL \footnotesize{$\uparrow$} & WER \footnotesize{$\downarrow$} & COS \footnotesize{$\uparrow$} & ~ \\ \midrule
        Noisy & 2.62 & 3.97 & 2.56 & 2.97 & 43.0 & - & 2.65 & 3.94 & 1.99 & 2.78 & 50.3 & - & ~ \\
        DCCRNet \cite{hu2020dccrn} & 4.15 & 4.08 & 4.26 & 3.73 & 21.6 & 0.89 & 3.22 & 4.10 & 3.66 & 3.26 & \textbf{30.8} & \underline{0.74} & ~ \\ 
        DCUNet \cite{choi2018phase} & \underline{4.20} & 4.14 & 4.20 & 3.71 & 19.5 & \textbf{0.91} & 2.97 & \textbf{4.27} & \underline{3.83} & \underline{3.47} & 33.5 & \textbf{0.75} & ~ \\
        DPTNet \cite{chen2020dual} & 4.12 & \underline{4.19} & 4.27 & \underline{3.78} & \textbf{17.6} & 0.91 & 3.11 & \underline{4.16} & 3.08 & 3.18 & 34.2 & 0.73 & ~ \\ 
        Conv-TasNet \cite{luo2019conv} & 4.12 & 4.18 & \underline{4.30} & \underline{3.78} & 18.2 & 0.91 & 3.23 & 4.03 & 3.06 & 3.11 & 33.9 & \underline{0.74} & ~ \\
        AnCoGen (ours) & \textbf{4.24} & \textbf{4.21} & \textbf{4.32} & \textbf{3.81} & \underline{18.0} & 0.73 & \textbf{4.18} & 4.01 & \textbf{4.20} & \textbf{3.48} & \underline{31.9} & 0.71 & ~ \\
        \bottomrule
    \end{tabular}
    }
    \captionof{table}{Speech denoising results (best score in each column is in bold, second best score is underlined).}
    \label{tab:speech-enhancement}
\end{figure*}

\subsection{Experimental setup}
\label{subsec-exp-setup}

\myparagraph{Tasks} In this work, AnCoGen is evaluated across four tasks: Analysis-resynthesis of clean speech signals, robust $f_0$ estimation, pitch shifting, and speech denoising. As illustrated in Fig.~\ref{fig:AnCoGen_high_level}, analysis-resynthesis consists of using AnCoGen to map an MS to the corresponding SAs (analysis) and then back to the MS (resynthesis), whereas pitch shifting and speech denoising consist of modifying the latent speech attributes ($A_2$ and $A_5$, respectively) before resynthesis. $f_0$ estimation is simply a byproduct of the analysis stage. Notice that AnCoGen could be evaluated on other tasks, such as voice conversion and speech dereverberation, which is left for future work (preliminary qualitative results are provided on the companion website\footref{note1}).\\

\myparagraph{Datasets} For the analysis-resynthesis, robust $f_0$ estimation and pitch shifting tasks, AnCoGen is trained on synthetic noisy and reverberant speech data. The signals are created using the 100-hour clean speech training subset of the LibriSpeech dataset \cite{panayotov2015librispeech}, which involves 251 different speakers, and the noise signals from the DEMAND dataset \cite{thiemann2013demand} (\textit{NRIVER, NPARK, TMETRO, TBUS, OOFFICE} noise types). The signal-to-noise ratio (SNR) was varied between $0$ and $30$ dB, and synthetic reverberation was added using \cite{sobot2021Pedalboard}. Evaluation data will be described individually for the three tasks mentioned above in the next subsection. For the denoising task, AnCoGen and the reference methods are trained and evaluated (matched conditions) on the single-speaker version of the LibriMix dataset \cite{cosentino2020librimix}, which is derived from LibriSpeech clean utterances \cite{panayotov2015librispeech} and WHAM! noises \cite{wichern2019wham}. To evaluate the methods on mismatched noise conditions, we also compute the performance on the LibriSpeech + DEMAND dataset, in which LibriSpeech test utterances are mixed with noise signals from DEMAND at an SNR of $0$~dB, using the noise types \textit{DWASHING, OHALLWAY, NFIELD, PCAFETER}, and \textit{OMEETING}.\\

\myparagraph{Settings} The Mel-spectrograms are computed using the short-time Fourier transform with a 64-ms Hann window (1024 samples at 16 kHz) and a 10-ms hop size (160 samples). The number of Mel bands is set to $128$. The model is trained for 800 epochs using the AdamW optimizer \cite{loshchilov2017decoupled} with a learning rate of $10^{-3}$. One epoch takes approximately 4 minutes on 4 A100 GPUs.\\

\myparagraph{Metrics} To measure the quality of the resynthesized/denoised speech signals, we use the following intrusive metrics: The SI-SDR (in dB) \cite{leroux2019sdr}, the wideband perceptual evaluation of speech quality (PESQ) measure \cite{rix2001perceptual}, the short-time intelligibility index (STOI) \cite{taal2011algorithm} and the cosine similarity (COS) between speaker voice embeddings obtained from Resemblyzer~\cite{wan2018generalized}. We also use the following non-intrusive metrics: NORESQA-MOS, a learning-based estimate of the subjective mean opinion score (MOS) using non-matching references \cite{manocha22c_interspeech, kumar2023torchaudio} (hereinafter referred to as N-MOS), and DNSMOS P.835 \cite{reddy2022dnsmos}, which provides an estimate of the speech signal quality (SIG), background noise intrusiveness (BAK), and overall quality (OVRL). For all these metrics, a higher value indicates a better result. We also compute the word error rate (WER, in $\%$) obtained from a pre-trained speech-to-text model\footnote{\url{https://huggingface.co/facebook/wav2vec2-base-960h}} applied on the reconstructed waveforms.

\subsection{Results and discussion}

\myparagraph{Analysis-resynthesis} 
Table~\ref{tab:quality-reconstruction} presents the analysis-resynthesis results, obtained using clean speech utterances from the LibriSpeech dataset. The results distinguish between cases where the speakers are seen during training and where they are unseen. The table compares the performance of the proposed method with that of HiFi-GAN when directly fed with the ground-truth MS (``GT MS"). The ``GT waveform" entry corresponds to the ground-truth clean speech waveform and indicates the best possible values. It can be seen that HiFi-GAN performs poorly on intrusive metrics when using the ground-truth MS. This is due to a poor phase estimation by this model, which causes misalignment with the target waveform, significantly affecting metrics like SI-SDR. However, subjective evaluations in \cite{kong2020hifi} showed minimal impact on perceived reconstruction quality. In the following sections, we will focus on the non-intrusive metrics. The GT MS and AnCoGen achieve high N-MOS values, close to those of the GT waveform and indicating good audio quality. The WER is slightly higher for AnCoGen (3.03 and 5.30\%) compared to the GT MS (2.28 and 2.31\%), but intelligibility remains satisfactory. However, the COS metric for AnCoGen drops significantly (from 0.90 to 0.75) from the seen to unseen speaker cases. This shows that AnCoGen struggles to generalize to unseen speakers, which was actually expected. This is due to the quantization of the speaker identity attribute, which forces AnCoGen to output the voice of one of the 251 speakers used for training.\\

\myparagraph{Robust $f_0$ estimation} 
For the robust $f_0$ estimation experiment, we mixed the clean speech signals of the PTDB-TUG database \cite{pirker2012pitch} with cafeteria noise from the DEMAND dataset (PCAFETER noise type, unmatched with the training conditions) \cite{thiemann2013demand}, varying the SNR from 0 to 10 dB. The $f_0$ ground-truth is obtained from laryngograph signals. We evaluate the performance with and without reverberation. The proposed approach is compared to the following reference methods: (i) pYin \cite{mauch2014pyin}, an enhanced version of the Yin algorithm using Viterbi decoding; (ii) SWIPE \cite{camacho2008sawtooth}, which is based on frequency-domain autocorrelation (we used the \texttt{pysptk} implementation\footnote{\url{https://github.com/r9y9/pysptk}}); and (iii) CREPE \cite{kim2018crepe}, which  we already used to generate training data (see Section~\ref{subsec:speech_rep}). The results are shown in Fig.~\ref{fig:pitch_estimation} regarding average absolute error (AAE, lower is better) for different SNRs and with or without reverberation. AnCoGen shows very accurate $f_0$ estimation performance, matching that of CREPE on clean data, and maintaining very good accuracy in noisy and reverberant conditions, with an AAE that remains lower than 6.7 Hz in all conditions. This strongly contrasts with pYin and SWIPE, whose accuracy significantly degrades with the SNR.\\

\myparagraph{Pitch shifting}
In this experiment, clean speech signals are manipulated by shifting the original $f_0$ trajectory by $\pm$10\% and $\pm$50\% in the control stage of AnCoGen (see Fig.~\ref{fig:AnCoGen_high_level}). Using again the PTDB-TUG database, we evaluate the performance in terms of AAE between the desired and predicted $f_0$ (using CREPE), and in terms of speech signal quality using again the N-MOS metric of \cite{kumar2023torchaudio}. We compare our approach to the WORLD vocoder \cite{morise2016world} and to the time-domain PSOLA algorithm \cite{moulines1990pitch}. The results in Table~\ref{tab:pitch-manipulation} show that AnCoGen is competitive with these reference methods. It achieves accurate $f_0$ control with a low AAE and maintains good speech quality with a high N-MOS, even for large shifts ($\pm$50\%). \\

\myparagraph{Speech denoising} In this experiment, noisy speech signals are denoised with AnCoGen by setting the SNR attribute $A_5$ to 40~dB in the control stage of AnCoGen. The proposed approach is compared to the deep complex convolution recurrent network (DCCRNet) of \cite{hu2020dccrn}, the deep complex U-net (DCUNet) of \cite{choi2018phase}, the dual-path transformer network (DPTNet) of \cite{chen2020dual}, and the fully-convolutional time-domain audio separation network (Conv-TasNet) of \cite{luo2019conv}. 
The results are presented in Table~\ref{tab:speech-enhancement}. AnCoGen consistently outperforms other methods in most signal quality-related metrics (N-MOS, SIG, BAK, OVRL) under both matched and mismatched conditions, while achieving the second-best performance in terms of WER. This demonstrates that AnCoGen effectively balances noise reduction, speech quality, and intelligibility. However, it ranks last in speaker identity preservation (as measured by COS), which is consistent with the analysis-resynthesis results discussed above. Nevertheless, some researchers argue that it may be desirable to prioritize overall quality over fidelity to the reference signal, such as in terms of speaker identity  \cite{hsu2023revise}. One notable strength of AnCoGen is its capacity to generalize its performance in noise suppression and speech quality when the test conditions differ from the training ones. AnCoGen maintains a very stable N-MOS and BAK performance between matched and mismatched noise conditions (only $-0.06$ points of N-MOS and $-0.12$ points of BAK), whereas the reference methods show a significant drop in performance (between $-0.89$ and $-1.23$ points of N-MOS and between $-0.37$ and $-1.24$ points of BAK). The high quality of denoised signals is ensured by the fact that AnCoGen is a generative model trained to generate clean signals. One drawback, though, is that AnCoGen sometimes outputs a denoised signal with a phonetic content that does not exactly correspond to the input noisy signal.

\section{Conclusion}

In this article, we presented AnCoGen, a bidirectional model that maps an audio signal into controllable speech attributes (such as pitch, loudness, identity, noise, and clarity index) and generates the corresponding signal from those attributes. 
Combined with a neural vocoder, the proposed approach allows for the analysis, control, and generation of speech signals using a single model.
Extensive experiments showed that AnCoGen accurately estimates pitch even under noisy conditions, enables precise pitch manipulation, and improves speech quality in background noise, even under mismatched conditions.
The main limitation of AnCoGen is its ability to recreate a speaker's voice for an unseen speaker accurately. Finer quantization of speaker embeddings could improve the voice reconstruction for new speakers, such as using a hierarchical codebook \cite{razavi2019generating}.

\bibliographystyle{IEEEtran}
\bibliography{refs}

\end{document}